\documentclass[aps,twocolumn,preprintnumbers,final]{revtex4}
\usepackage{graphicx,amsmath,amssymb,mathrsfs,array,longtable,delarray,verbatim,epsfig}
\graphicspath{{E:\My_papers\Multiplexer}}
\begin{document}
\title{The effect of split gate size on the electrostatic potential and 0.7 anomaly within one-dimensional quantum wires on a modulation doped GaAs/AlGaAs heterostructure}
\author{L. W. Smith$^{1,\dagger,*}$, H. Al-Taie$^{1,2}$, A. A. J. Lesage$^{1}$, K. J. Thomas$^{3}$, F. Sfigakis$^{1}$, P. See$^{4}$, J. P. Griffiths$^{1}$, I. Farrer$^{1,\ddagger}$, G. A. C. Jones$^{1}$, D. A. Ritchie${^1}$, M. J. Kelly$^{1,2}$, and C. G. Smith$^{1}$}
\affiliation{
$^{1}$Cavendish Laboratory, Department of Physics, University of Cambridge, J. J. Thomson Avenue, Cambridge, CB3 0HE, United Kingdom\\
$^{2}$Centre for Advanced Photonics and Electronics, Electrical Engineering Division, Department of Engineering, 9 J. J. Thomson Avenue, University of Cambridge, Cambridge CB3 0FA, United Kingdom\\
$^{3}$Department of Electronic and Electrical Engineering, University College London, Torrington Place, London WC1E 7JE, United Kingdom\\
$^{4}$National Physical Laboratory, Hampton Road, Teddington, Middlesex TW11 0LW, United Kingdom}
\date{\today}
           
\begin{abstract}
We study 95 split gates of different size on a single chip using a multiplexing technique. Each split gate defines a one-dimensional channel on a modulation-doped GaAs/AlGaAs heterostructure, through which the conductance is quantized. The yield of devices showing good quantization decreases rapidly as the length of the split gates increases. However, for the subset of devices showing good quantization, there is no correlation between the electrostatic length of the one dimensional channel (estimated using a saddle point model), and the gate length. 
The variation in electrostatic length and the one-dimensional subband spacing for devices of the same gate length exceeds the variation in the average values between devices of different length.
There is a clear correlation between the curvature of the potential barrier in the transport direction and the strength of the ``0.7 anomaly'': the conductance value of the 0.7 anomaly reduces as the barrier curvature becomes shallower. These results highlight the key role of the electrostatic environment in one-dimensional systems. Even in devices with clean conductance plateaus, random fluctuations in the background potential are crucial in determining the potential landscape in the active device area such that nominally identical gate structures have different characteristics.
\end{abstract}
\maketitle
\section{Introduction} 
Low dimensional systems are platforms for research into a variety of quantum behavior. However, the broader application of such devices in an engineering context is limited due to unpredictable local variations in the electrostatic landscape in the device's active area. When considering their practical application it is important to study the statistical spread in device characteristics. We have previously studied identical devices fabricated on a modulation doped heterostructure which show variations in their conductance properties~\cite{Smith2014}. In this paper we investigate the effect of gate size on two important quantum properties of split gate devices--the conductance quantization \cite{Wharam1988, vanWees1988} and the occurrence of 0.7 anomaly \cite{Thomas1996}.

The split gate is the simplest mesoscopic device that can be used to study how device behavior is affected by gate size.
The conductance through a split gate~\cite{Thornton1986} is quantized in multiples of $G_0=2e^2/h$ as a function of the voltage applied to the gates~\cite{Wharam1988, vanWees1988}, due to the formation of one-dimensional (1D) subbands. For an ideal 1D conductor, this does not depend on the length as long as transport remains ballistic. 
The effect of split gate size can be investigated either by varying lithographic dimensions \cite{Lee2006, Koop2007, Reilly2001, Koester1996}, or fabricating several split gates in close-proximity which act in series to modify the potential of a single 1D channel~\cite{Thomas2004, Iqbal2013, Iqbal2013b, Heyder2014}.
So far it has been shown that the split-gate voltage ($V_{sg}$) at which the conductance through the 1D channel is pinched off occurs closer to zero for longer and narrower devices~\cite{Lee2006, Koop2007}. 
Additionally, the quality of conductance quantization degrades as the gate length increases \cite{Timp1989, Koester1996}. This latter effect is related to the higher probability of encountering an impurity in the channel with longer/wider split gates, and fluctuations in the background disorder potential which modifies the potential landscape in the channel area.

Some studies of the effect of split gate size have focused on the 0.7 anomaly \cite{Koop2007, Reilly2001, Iqbal2013}, a conductance feature which occurs near $0.7G_0$~\cite{Thomas1996, Thomas1998, Micolich2011} which arises from enhanced electron interactions at low conductance. A reduction in the conductance of the 0.7 anomaly for longer split gates has been reported~\cite{Reilly2001}. Another study using split gates in series has shown periodic modulations in the value of the 0.7 anomaly as a function of 1D channel length \cite{Iqbal2013}.
The origin of the 0.7 anomaly is currently debated, theories proposed for its occurrence include spontaneous spin polarization~\cite{Thomas1996, Wang1996}, the Kondo effect~\cite{Cronenwett2002, Meir2002, Rejec2006, Iqbal2013}, Wigner crystallization~\cite{Matveev2004, Brun2014}, and inelastic scattering plus the local enhancement (smeared van Hove singularity) of the 1D density of states~\cite{Sloggett2008, Bauer2013}. 

We use a multiplexing scheme~\cite{Al-Taie2013, Smith2014, Al-Taie2015, Lesage2015, Smith2015, Puddy2015} to measure 95 split gates of 7 different sizes on a GaAs/AlGaAs modulation-doped heterostructure. 
This is the first study of the impact of split gate dimensions using data obtained from a large number of split gates fabricated on a single chip and measured during one cool down. The array contains groups of devices, where the dimensions of each split gate in the group are the same, and dimensions are changed between groups. Measuring groups in this way provides enough statistical information to determine typical device properties. 

Our technique allows us to compare incremental changes in gate design. This is important in the context of device developement, for example if it is necessary to design a device with a specific set of operating parameters, or find ways of improving a particular design. 
The traditional approach to nanostructure measurement require many cool downs to build up the statistical information that we obtain in a single cool down.

We systematically compare the accuracy of conductance quantization between split gates of different dimension, and find a rapid reduction in quality of quantization as gate length increases. 
We obtain average values of pinch off voltage and definition voltage as a function of gate length and width which show how the pinch off voltage occurs closer to zero for longer and narrower split gates, as expected from electrostatics \cite{Davies1995}. For longer gates the voltage at which the 1D channel is defined also occurs closer to zero, and the 1D subband spacing reduces. 

A key finding of this paper is that the background disorder potential is at least as important as the split gate dimension in governing the potential landscape in the 1D channel. This is shown in three separate ways. Firstly, the spread in both the measured 1D subband spacing and the estimated 1D barrier curvature for split gates at fixed gate lengths exceed variations in the mean value of these parameters between devices of different length. Secondly, the strength of coupling between the split gates and the 1D channel does not monotonically increase with device length. A monotonic increase is expected if one only considers the electric field generated by the gates themselves.
Thirdly, changes in the 1D barrier curvature from device to device--which indicates the length of the 1D channel--do not depend on split gate length.

We also directly compare the 0.7 anomaly in devices of different gate length. This is possible since--for a non-interacting system--the shape of the conductance trace depends on $\hbar \omega_{x,1}$. We use a technique developed in Ref.~\cite{Smith2015} to remove the trivial geometric dependence from the conductance data, leaving differences that are only due to electron interactions.
Our data are consistent with Ref.~\cite{Smith2015} in that the 0.7 anomaly occurs at lower conductance values for devices with a shallower longitudinal barrier. However, the present dataset is gathered from devices of different dimensions (data in Ref.~\cite{Smith2015} are from lithographically identical gates). By obtaining the same result from devices with various dimensions we show that the electrostatic length of the 1D channel is the decisive factor governing the conductance of the 0.7 anomaly.

\section{Methods}
\begin{figure}
\includegraphics[width=8cm,height=20cm,keepaspectratio]{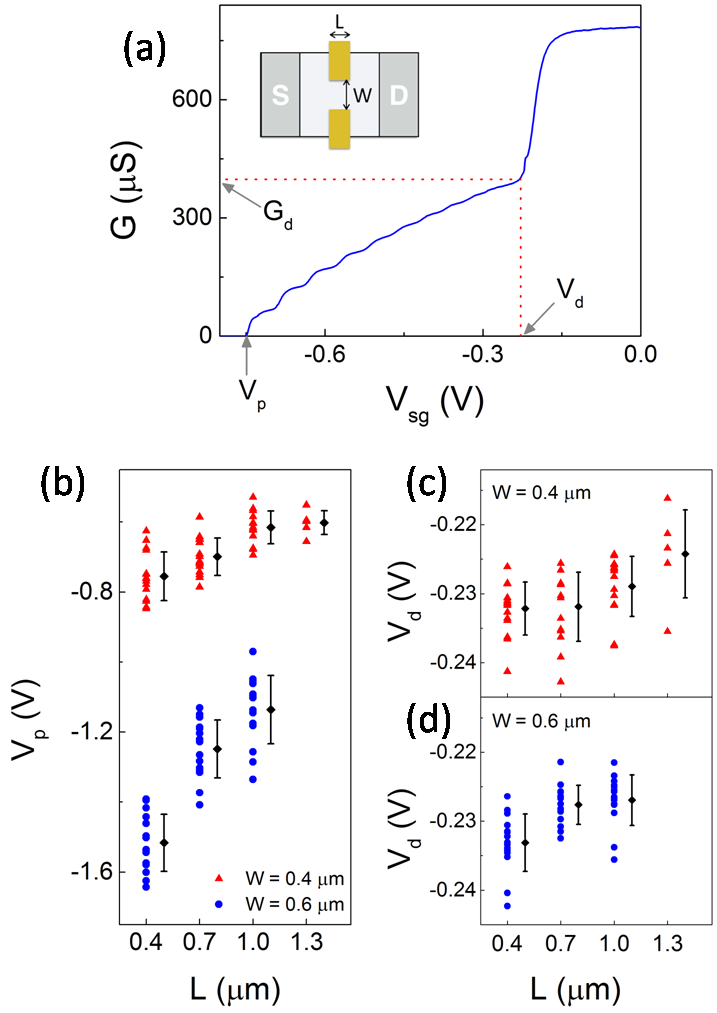}
\caption{\label{Fig1} 
(a) Typical trace of conductance $G$ as a function of the voltage applied to the split gate $V_{sg}$. The vertical and horizontal dotted lines indicate the 1D definition voltage ($V_d$) and conductance ($G_d$), respectively. The left-hand arrow indicates pinch-off voltage $V_p$. The inset shows a schematic split gate device with dimensions $L$ and $W$ labeled. Source and drain ohmic contacts are marked S and D, respectively.
(b) Scatter plot of pinch-off voltage as a function of split-gate length $L$.
The triangles (circles) represent data from devices of width $W=0.4$ ($0.6$) $\mu$m. The diamonds and error bars show the mean and standard deviation for each $L$, respectively, offset horizontally by $0.1$ $\mu$m for clarity.
Panels (c) and (d) show $V_d$ as a function of $L$, for $W=0.4$ and $0.6$ $\mu$m, respectively. }
\end{figure}

Our data are obtained from 95 split gates of various dimensions located within a single array (of total area $1.5 \times 1.95$ mm$^2$).
Two widths were chosen, $W=0.4$ and $0.6$ $\mu$m [the dimensions are indicated on a schematic diagram of a split gate in Fig.~\ref{Fig1}(a)].
For both widths, 15 devices of each length $L$ were measured, where $L = 0.4$, $0.7$, and $1.0$ $\mu$m. One of the devices did not define a 1D channel ($L/W = 1.0/0.4$ $\mu$m). A further 6 devices with $L/W = 1.3/0.4$ $\mu$m were also measured.

The sample is fabricated on a GaAs/AlGaAs heterostructure in which the 2DEG forms 90 nm below the surface of the wafer, separated from a 40 nm wide Si-doped AlGaAs layer by 40 nm of undoped AlGaAs.
The carrier density and mobility are $2.0\times10^{11}$ cm$^{-2}$ and $3.0\times10^6$ cm$^2$V$^{-1}$s$^{-1}$, respectively, determined by measuring a Hall bar on a nearby piece of the wafer. The electron mean free path is 22 $\mu$m.
%The phase coherence length is calculated to be 52 nm~\cite{Beenakker1991}.
%W0476

The split gates are defined by electron-beam lithography. Other surface gates are defined by optical lithography, and all gates are metallized by thermally evaporating Ti/Au. 
A two-terminal, constant voltage technique is used to measure the differential conductance through the split gates as a function of gate voltage, using an ac excitation voltage of 100 $\mu$V at 17 Hz.
All measurements are carried out at 1.4 K and $B=0$ T.

The device measured here is different to that used in our previous work on multiplexed split gates~\cite{Al-Taie2013, Smith2014, Al-Taie2015, Smith2015}. The earlier work used an array of identical split gates, where both the length and width were 0.4 $\mu$m. The multiplexed array measured here contains split gates of 7 different length/width combinations and is fabricated on a higher mobility heterostructure. This paper presents a self-contained story of the influence (or lack thereof) of gate length on conductance in 1D devices. We investigate the effect of split gate dimensions on device yield, 1D conductance properties, the electrostatic potential profile and the 0.7 anomaly. Such a systematic study on many split gates with different dimensions and measured under identical conditions is the first of its kind.

 % saved Desktop 31-10-14\10Tesla2014\forPaper\OriginPlots.opj
\begin{figure}
\includegraphics[width=9cm,height=6cm,keepaspectratio]{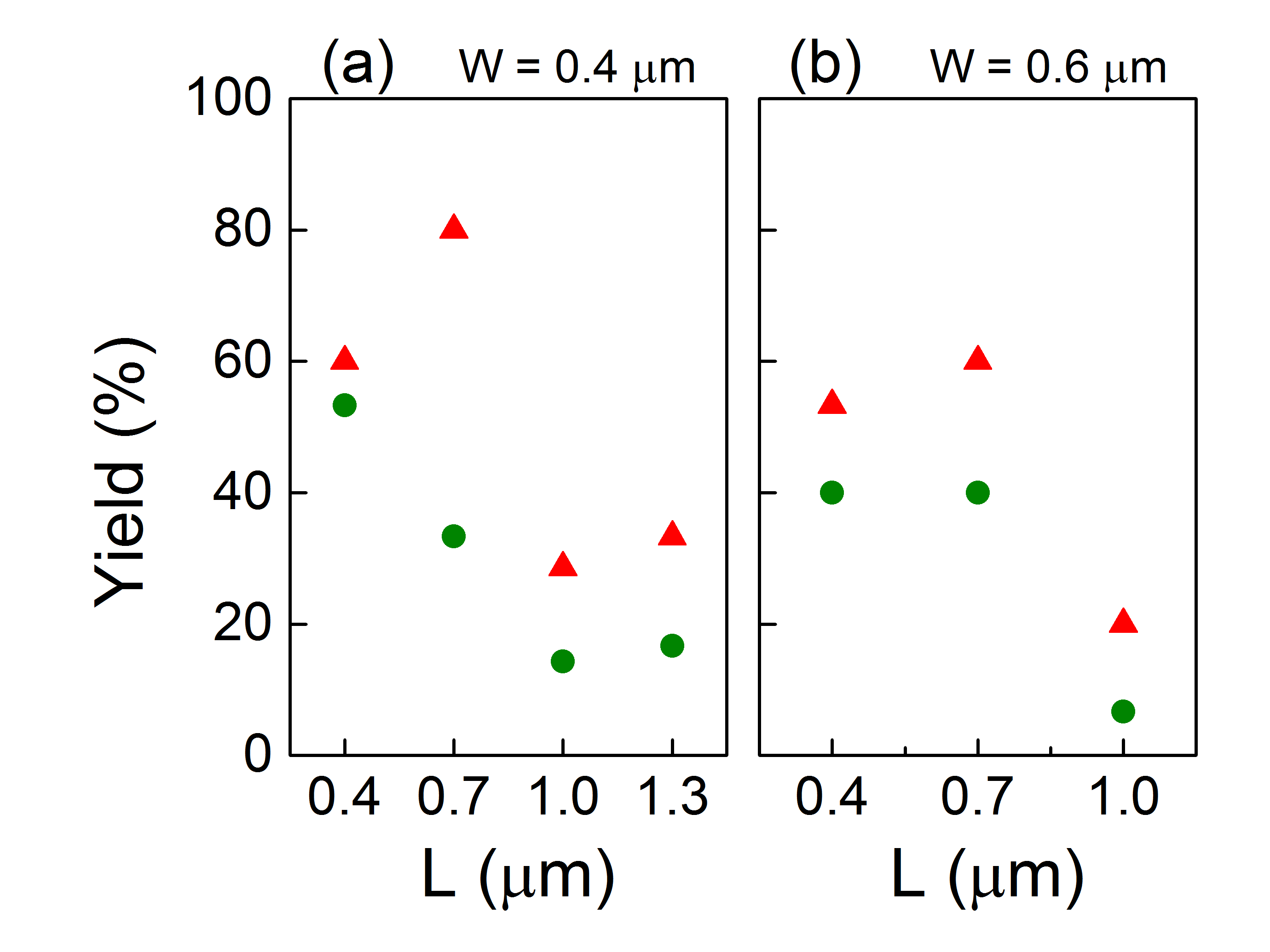}
\caption{\label{Fig2} Percentage of devices showing clean quantization as a function of split-gate length $L$. Panels (a) and (b) show data for widths $W=0.4$ and $0.6$ $\mu$m, respectively. The triangles (circles) indicate devices for which the first and second (first, second, and third) plateaus occur within $\pm0.1G_0$, after correcting for series resistance.
}
\end{figure} % saved Desktop 31-10-14\10Tesla2014\forPaper\OriginPlots.opj
\section{Conductance properties as a function of split gate size}
\subsection{Electrical properties}
We begin by investigating the effect of split gate length and width on the definition voltage $V_d$ and pinch-off voltage $V_p$. 
Figure~\ref{Fig1}(a) shows a typical plot of the conductance as a function of the voltage applied to the split gate $V_{sg}$. A 1D channel is formed when the 2DEG beneath the gates is fully depleted, indicated by a sudden change in the gradient of the conductance trace [corresponding to voltage $V_d$ and conductance $G_d$ in Fig.~\ref{Fig1}(a)]. 
As $V_{sg}$ decreases further the conductance reduces (showing a series of plateaus) until the channel is completely pinched off (marked by $V_p$).

Figure~\ref{Fig1}(b) shows a scatter plot of $V_p$ against $L$, in which the triangles (circles) represent data for $W=0.4$ ($0.6$) $\mu$m. The diamonds/error bars show average values/standard deviation for each $L$. 
Both sets of data show $V_p$ becoming closer to zero with increasing $L$ \cite{Koop2007, Lee2006}. The trend is more pronounced for $W=0.6$ $\mu$m, where $|V_p|$ is nearly double that of $W=0.4$ $\mu$m. These trends arise from simple electrostatics, since for a given $V_{sg}$, the electric field is stronger in the center of the channel for longer and narrower split gates \cite{Davies1995}.
An additional effect also occurs: a longer wire is more likely to be affected by fluctuations in the background disorder potential due to ionized donors. This can modify the confining potential and therefore $V_p$. 
The larger spread in $V_p$ for $W=0.6$ $\mu$m may reflect the increased role of disorder. However, the spread as a percentage of the mean is similar for both $W$. 

Figures~\ref{Fig1}(c) and (d) show $V_d$ against $L$ for $W=0.4$ and $0.6$ $\mu$m devices, respectively. 
In both cases the magnitude of $V_d$ reduces as $L$ increases, which can be attributed to the higher electric field strength in the center of the channel for longer devices.
For each $L$, the range and average values of $V_d$ are similar for both widths. 

\subsection{Yield}
We investigate the role of disorder as a function of device length/width.
The values of conductance plateaus are used to define a yield criterion, since when disorder affects the transmission through a 1D channel the conductance plateaus deviate from expected values.
For systematic analysis the data are first corrected for series resistance ($R_s$) using $R_s=1/G$ at $V_{sg}=0$ V (i.e. the open-channel resistance).
Two cases (A and B) are considered. Case A follows Ref.~\cite{Al-Taie2013} and requires the first two conductance plateaus to occur within $\pm 0.1G_0$. Case B extends this to include the third plateau.

Figures~\ref{Fig2}(a) and~\ref{Fig2}(b) show the number of devices which pass the yield criterion as a function of $L$, for $W=0.4$ and $0.6$ $\mu$m, respectively.
Triangles and circles indicate devices which passed for cases A and B, respectively.
Fewer devices show accurate conductance quantization as $W$ or $L$ increases~\cite{Timp1989, Koester1996}, due to the increased likelihood of encountering impurities in the 1D channel and greater variation in the background potential. 
Including the third plateau in the analysis gives a stricter yield criterion and leads to a lower yield.

The specific form of the relationship between yield and length in Fig. \ref{Fig2} is not clear: it is necessary to measure more than 15 devices of each $L$ to obtain this information. This is an interesting avenue for future research since finding the exact nature of the correlation between yield and length may provide information about disorder correlation lengths or dominant disorder effects in the 1D channel. 

\begin{figure}
\includegraphics[width=16cm,height=12cm,keepaspectratio]{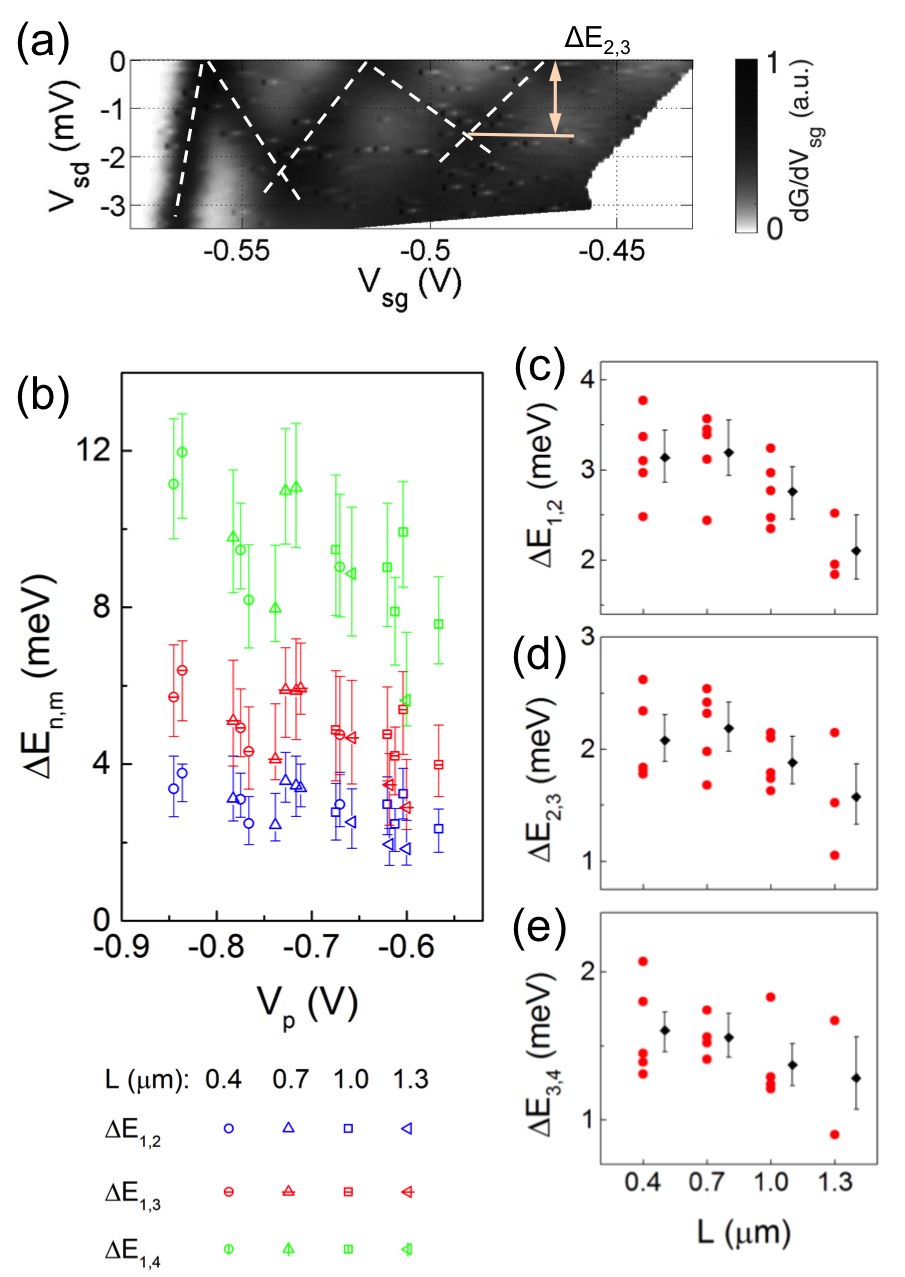}
\caption{\label{Fig3} (a) Grayscale diagram of the transconductance $dG/dV_{sg}$ as a function of $V_{sg}$ and source-drain bias $V_{sd}$, from a typical device with dimensions $L/W = 0.4/0.4$. Black (white) regions correspond to high (low) transconductance, i.e. transitions between plateaus (the plateaus themselves). The 1D subband spacings are estimated by the crossings of peaks in the transconductance \cite{Patel1991}, highlighted by dashed lines. Subband spacing $\Delta E_{2,3}$ is labeled for illustrative purposes.
(b) Cumulative 1D subband spacings $\Delta E_{n,m}$ as a function of pinch-off voltage $V_p$. The blue, red, and green symbols correspond to $\Delta E_{1,2}$, $\Delta E_{1,3}$, and $\Delta E_{1,4}$, respectively. Unique symbols represent devices of each size, as described in the legend. The error bounds show the cumulative error in the estimate.
(c)-(e) Scatter plots of $\Delta E_{n,n+1}$ against $L$ for $n=1$, $2$ and $3$, respectively (to avoid confusion the error bars on individual data points are not shown). The diamonds show the average for each $L$ (offset horizontally by 0.1 $\mu$m for clarity), and error bounds indicate the average error.
}
\end{figure} % saved Desktop 31-10-14\10Tesla2014\forPaper\OriginPlots.opj
\section{Dependence of the electrostatic potential landscape on gate length}
Having characterized the length- and width-dependence of two electrical properties of the 1D conductance trace ($V_p$ and $V_d$), we now consider the length-dependence of the potential profile for a subset of devices.
We first investigate the transverse confining potential by extracting the 1D subband spacing \cite{Patel1991}. We then study the longitudinal potential by estimating the curvature of the barrier in the direction of electron transport \cite{Smith2014}.
The following measurements and analysis are performed using devices with $W=0.4$ $\mu$m, since a higher percentage of split gates with $W=0.6$ $\mu$m were affected by disorder.
\subsection{DC bias spectroscopy}
DC bias spectroscopy is used to measure the 1D subband spacing~\cite{Patel1991} for 18 split gates.
Figure~\ref{Fig3}(a) shows an example grayscale plot of the transconductance $dG/dV_{sg}$ as a function of source-drain bias $V_{sd}$ and $V_{sg}$ (dark/light regions correspond to high/low transconductance).
The data are corrected for series resistance following the procedure outlined in the supplementary material of Ref.~\cite{Srinivasan2013}.

As $| {V_{sd}} |$ increases, dark regions representing peaks in the transconductance diverge into two (highlighted by the dashed lines), corresponding to the bottom of the 1D subband reaching either the source or drain chemical potential~\cite{Thomas1995}. 
Two lines cross when $V_{sd}$ is equal to the energy difference between consecutive 1D subbands, giving the 1D subband spacing $\Delta E_{n,n+1}$ ($n$ is the subband index). For example, $\Delta E_{2,3}$ is marked in Fig.~\ref{Fig3}(a).

Figure~\ref{Fig3}(b) shows cumulative 1D subband spacings $\Delta E_{n,m}$ as a function of $V_p$ for 18 devices.
The spacing between the first and second ($\Delta E_{1,2}$), first and third ($\Delta E_{1,3}$), and first and fourth ($\Delta E_{1,4}$) subbands are shown (unique symbols represent devices of different $L$, described in the legend).

In each case $\Delta E_{n,m}$ reduces as $V_p$ becomes closer to zero.
We plot the cumulative data to better accentuate this trend, although 
the spacing between consecutive subbands [shown in Figs. \ref{Fig3}(c)--(e)], all show a downward trend with $V_p$.
The trend can be attributed to the weakening of the electric field at pinch off for smaller $|V_p|$, since this leads to shallower confinement and therefore closer subband spacings.

Figures~\ref{Fig3}(c)-(e) show individual 1D subband spacings as a function of $L$. Panels (c), (d), and (e) show $\Delta E_{1,2}$, $\Delta E_{2,3}$, and $\Delta E_{3,4}$, respectively. 
The diamonds represent average values for each $L$ (offset horizontally for clarity), and error bounds show the average error for each $L$~\cite{NoteError}.
The average subband spacing reduces with $L$ in agreement with electrostatic modeling of a saddle point potential~\cite{Koop2007, Davies1995}. 
However, the spread in $\Delta E_{n,n+1}$ for individual values of $L$ is larger than or similar to the average change in $\Delta E_{n,n+1}$ across all values of $L$. 
Since variations in device characteristics likely arise from fluctuations in the background potential, these data highlight the importance of unique electrostatic environment near each split gate. They suggest that backgound variations are as significant as the lithographic dimensions in governing the potential landscape in the device active area.

\subsection{Coupling between the split gate and the 1D channel}
\begin{figure}
\includegraphics[width=8.5cm,height=20cm,keepaspectratio]{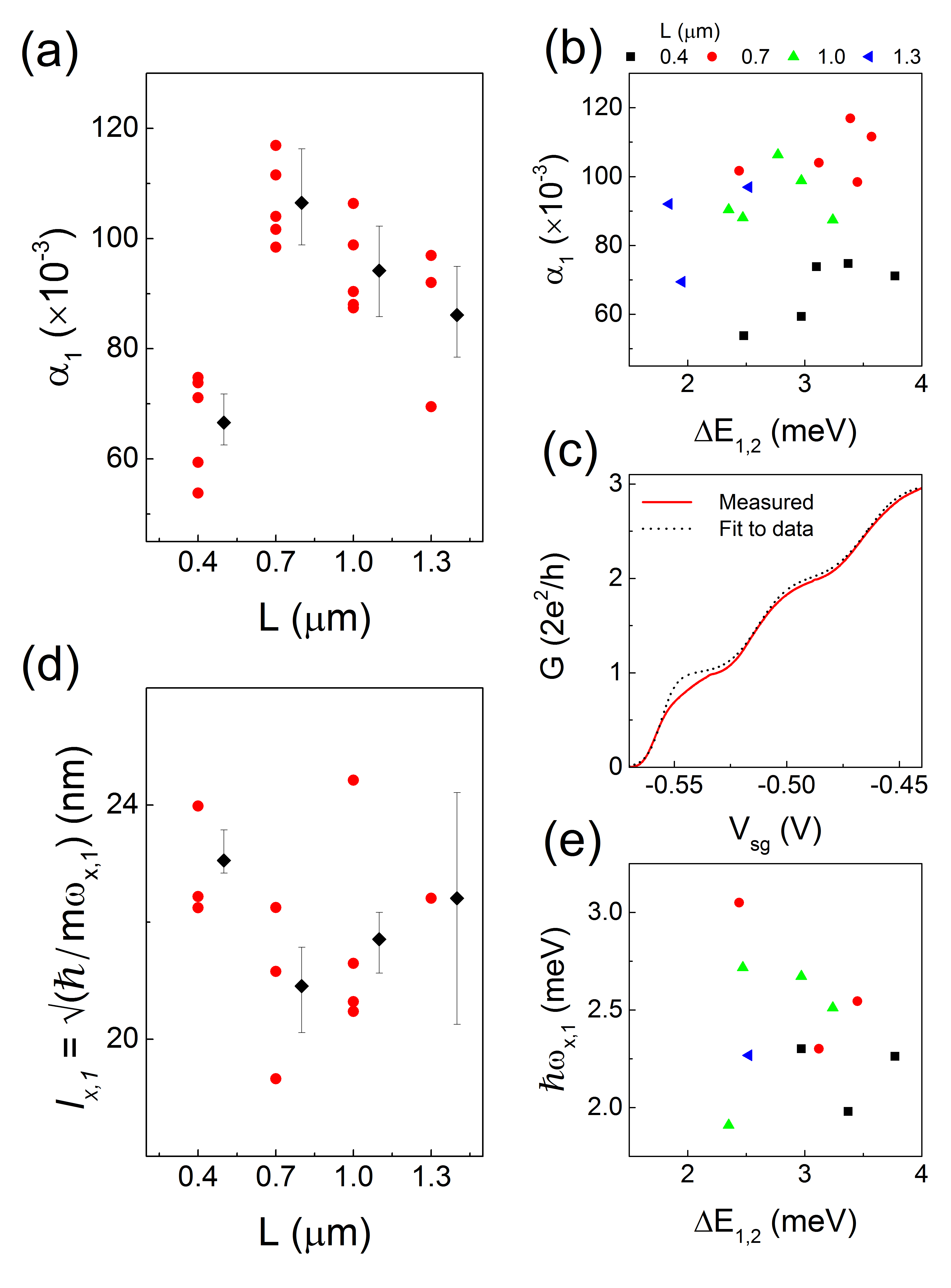}
\caption{\label{Fig4}
(a), (b) Lever arm $\alpha_1$ as a function of lithographic length $L$ and 1D subband spacing $\Delta E_{1,2}$, respectively. The diamonds in panel (a) represent the mean for each $L$ (offset by 0.1 $\mu$m for clarity), and the error bars indicate the average error.
(c) Experimentally measured conductance $G$ as a function of split-gate voltage $V_{sg}$, for an example device (solid line). The dashed line shows a fit to the data using a modified saddle-point model. The transition between plateaus gives an estimate of barrier curvature $\hbar\omega_{x,n}$. 
(d) Characteristic length of the potential barrier $\sqrt{\hbar / m^*\omega_{x,1}}$ as a function of gate length $L$. The diamonds and error bars represent the average value and the average error, respectively (offset by 0.1 $\mu$m for clarity).
(e) Curvature of the potential barrier $\hbar\omega_{x,1}$ as a function of $\Delta E_{1,2}$.
}
\end{figure}
The degree of coupling between the split gate and the 1D channel is given by lever arm $\alpha = \Delta E /e\Delta V_{sg}$, where $\alpha=\partial V_{sd}/ \partial V_{sg}$ (estimated from DC bias spectroscopy measurements). Figure~\ref{Fig4}(a) shows $\alpha_1$ (the lever arm for the first subband), as a function of $L$ for 18 split gates. The diamonds indicate the average value, offset horizontally for clarity. The error bounds indicate the average error~\cite{NoteError}. 

The lever arm $\alpha_1$ almost doubles between $L=0.4$ and $0.7$ $\mu$m. 
This trend is expected since $\alpha$ scales with $\Delta E$, which, for a given $\Delta V_{sg}$, will be larger in the center of a channel formed by a longer gate.
However, as $L$ increases further, $\alpha_1$ reduces, deviating from the expected trend. This also occurs for subbands 2 and 3.

A length of $0.7$ $\mu$m is consistent with the correlation length of fluctuations in the background potential. 
In Ref.~\cite{Nixon1991}, calculations of the effect of potential fluctuations in the 2DEG on split gate devices are performed, for a heterostructure with a similar donor density to the one measured here \cite{noteNixon}. The fluctuations were assumed to arise from the random positioning of donors in the doped layer.
The correlation length of these fluctuations was found to be comparable with the device active area for split gate lengths of 0.2 and 0.6 $\mu$m.
Our data show $\alpha$ depends on split gate length up to $L = 0.7$ $\mu$m, which may suggest the correlation length of potential fluctuations is slightly greater than this value for the heterostructure used in our experiment.

Figure~\ref{Fig4}(b) shows $\alpha_1$ as a function of $\Delta E_{1,2}$. 
Data from each length split gate are represented using different symbols, described in the legend. Overall, no trend is apparent. Data from the same length devices appear to be grouped, and a weak positive correlation exists for the $L=0.4$ $\mu$m data. This does not occur for longer devices, consistent with the increased effect of disorder.

\subsection{Curvature of the 1D potential barrier}

We now estimate the curvature of the potential barrier in the transport direction $\hbar\omega_x$ by assuming the confining potential is described by a saddle-point model~\cite{Buttiker1990}.
We achieve this by fitting the measured data with a conductance calculated using the Landauer-B\"{u}ttiker formalism~\cite{Smith2014, Smith2015}, simulating a system of non-interacting electrons traversing a saddle-point potential with transmission probability $T_n = [1+\exp(-2\pi(E-E_n)/\hbar\omega_{x,n})]^{-1}$, where $E_n$ is the energy of the bottom of subband $n$.
The saddle-point approximation can be used for the first few subbands even for devices with a large length-to-width ratio, since $G$ is governed by transmission through the narrowest part of the channel.
 
Figure~\ref{Fig4}(c) shows measured conductance (solid line) and the fit (dashed line) for an example device, as a function of $V_{sg}$.
The fit is achieved as follows:
Transmission probability $T_n$ is calculated for the first three subbands individually as function of $E$, using an initial input of $\hbar\omega_{x,n} = \Delta E_{n,n+1}$. Subband-dependent lever arms $\alpha_n$ measured for each device individually are used to convert $E$ to a voltage scale. 
A minimization routine then optimizes $\hbar\omega_{x,n}$ to find the best fit between the calculated and measured traces. The use of subband-dependent values of $\hbar\omega_{x,n}$ reflects how the barrier profile for higher subbands is modified by the increased presence of electrons.
The sum of $T_n$ for $n=1$, 2 and 3 gives the final trace shown by the dashed line in Fig.~\ref{Fig4}(c).
 
The transition between $G=0$ and $0.5G_0$ is almost independent of temperature up to at least $T \approx 1.5$ K~\cite{Thomas1996}. 
Therefore, for $n=1$ the fit is performed with $T=0$, such that the calculated conductance $G_n=G_0 T_n$.
For higher subbands a temperature dependence is observed experimentally, therefore for $n=2$ and 3 we calculate $G_n$ at $T=1.4$ K using
\begin{equation}
G_n = G_0 \int\;dE\; \left(-\frac{\partial f}{\partial E}\right)\;T_n\ ,
\end{equation}
where $f$ is the Fermi-Dirac distribution.
We find that using either $T=0$ or $T=1.4$ for $n=1$ does not affect the trends observed~\cite{Smith2015}.
 
Good fits are obtained for 11 of the 18 split gates. For the other seven devices, conductance plateaus are weakened or suppressed due to strong disorder effects, therefore these data are discarded in the following analysis.
The saddle-point model assumes a parabolic potential barrier, which has a characteristic length $l_x = \sqrt {\hbar / m^*\omega_{x,n}}$, where $m^*$ is the effective mass of the electrons. Barrier length $l_x$ is therefore the distance over which the potential changes by $\hbar\omega_x/2$. 

Figure~\ref{Fig4}(d) shows $l_{x,1}$ as a function of $L$. Diamonds show the average $\bar{l}_x$ for each $L$, and error bounds indicate the average error, offset horizontally for clarity. We would expect the barrier length to scale with the gate length. However, our data show the opposite trend with $\bar{l}_x$ initially reducing for $L$ from $0.4$ to $0.7$ $\mu$m.
This change is very small compared to the change of $L$: $\Delta\bar{l}_x \simeq 2.5$ nm (a decrease of $11\%$), as $L$ increases by $175\%$.
As $L$ increases further, $\bar{l}_x $ then increases slightly.
For our devices therefore, the split gate length is not a good indicator of the length of the potential barrier.

For $L=0.4$, 0.7 and 1 $\mu$m, the spread of $l_{x,1}$ at fixed $L$ is larger than or similar to the overall change in average $l_{x,1}$ as $L$ varies.
This highlights the crucial importance of the background potential in determining the electrostatic landscape in the active device area, even between devices with nominally identical length.
This is further supported by Fig.~\ref{Fig4}(e), which shows $\hbar\omega_{x,1}$ as a function of $\Delta E_{1,2}$. For a given $L$, there is no correlation between these parameters. The trends for subbands 2 and 3 are similar. 
\begin{figure}
\includegraphics[width=16cm,height=11.5cm,keepaspectratio]{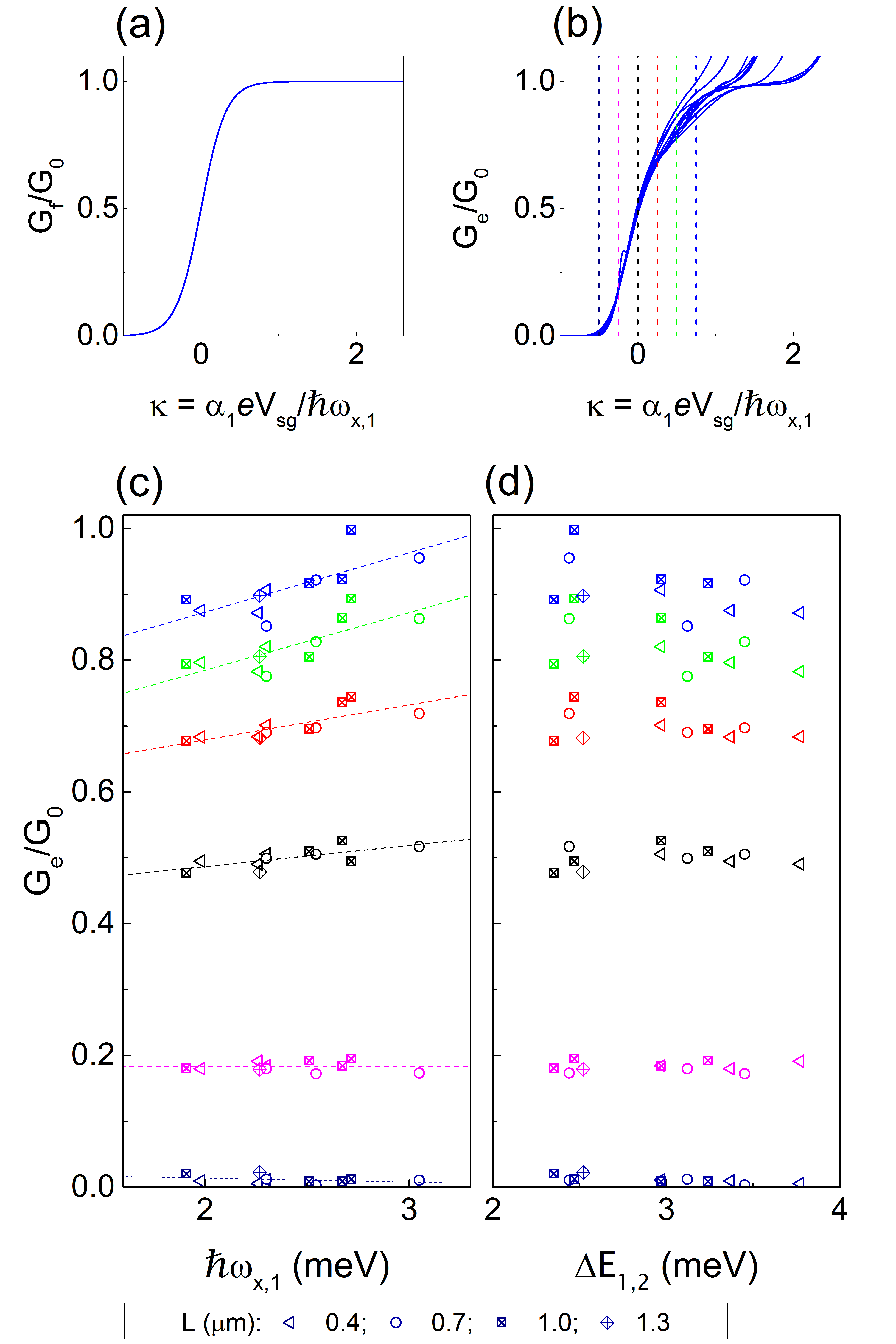}
\caption{\label{Fig5}
(a) Fitted conductance $G_f/G_0$  for the first 1D subband from 11 devices. Data are collapsed onto a universal curve by aligning $G_f/G_0=0.5$ with $V_{sg}=0$, then scaling $V_{sg}$ by $\alpha_1 e/\hbar\omega_{x,1}$.
(b) Corresponding experimentally-measured conductance $G_e/G_0$, where the data are offset and scaled using the same parameters as (a). For $G_e/G_0<0.5$ the traces collapse onto a similar curve. Above $0.5G_0$, variations arise due to the differences in the 0.7 anomaly. 
%The vertical lines show $\kappa=-0.5$ to $0.75$ in steps of $0.25$.
(c) $G_e/G_0$ as a function of barrier curvature $\hbar\omega_{x,1}$, at fixed values of the scaled voltage axis $\kappa$.
From bottom-to-top, $\kappa$ increases from $-0.5$ to $0.75$ in steps of $0.25$ [corresponding to vertical dashed lines, left-to-right, in panel (b)]. 
The dashed lines show a linear least squares fit to $G_e/G_0$ for each value of $\kappa$, as a guide to eye.
(d) Conductance $G_e/G_0$ as a function of spacing between the first and second 1D subbands $\Delta E_{1,2}$. 
In panels (c)--(d), data points for each length split gate are indicated by the symbols defined in the legend. }
\end{figure} % saved Dropbox 31_10_14\Multiplexer\Paper_length_variations\07Analysis_Fig5_alterations.opj - Fig2_T0
\section{0.7 anomaly in different length split gates}
In this section 0.7 anomalies from devices of different gate length are compared. Following Bauer \emph{et al.}~\cite{Bauer2013}, we refer to the conductance between $0.5G_0$ and $G_0$ as the `sub-open' regime. Additionally, we refer to experimentally measured (fitted) conductance data as $G_e$ ($G_f$).

Direct comparison of the 0.7 anomaly between devices is possible by removing the trivial -- that which can be accounted for in a non-interacting scenario \cite{Buttiker1990} -- dependence of
the conductance transition between $G=0$ and $G_0$ on barrier curvature.
This is achieved by offseting the conductance traces horizontally to align $G_f/G_0=0.5$ to $V_{sg}=0$ and scaling each $V_{sg}$ axis by $\alpha_1 e/\hbar\omega_{x,1}$ \cite{Smith2015}.
Differences in conductance that remain are only due to electron interactions. The strength of these interactions still depends on barrier shape~\cite{Sloggett2008, Bauer2013}.

Figure~\ref{Fig5}(a) shows the fitted conductance $G_f$ for the first subband as a function of the scaled voltage axis $\kappa = \alpha_1 e V_{sg}/\hbar\omega_{x,1}$. Data from all 11 split gates are plotted.
The traces collapse onto a universal curve since these data are obtained using a a non-interacting model. 
Figure~\ref{Fig5}(b) shows the corresponding $G_e/G_0$ data after applying the same scaling procedure.
For $\kappa \leq 0$, the traces collapse onto a very similar curve.
For $\kappa > 0$, differences occur due to the 0.7 anomaly (a spread in $G_e/G_0>1$ arises due to the different $\Delta E_{1,2}$).
The variation in the 0.7 anomaly between traces is related to the varying barrier curvature from device to device.

In order to compare the 0.7 anomaly between devices, $G_e/G_0$ is plotted as a function of $\hbar\omega_{x,1}$ in Fig.~\ref{Fig5}(c), for 6 fixed values of $\kappa$.
From bottom-to-top, $\kappa$ increases from -0.5 to 0.75 in steps of 0.25 [corresponding to the vertical lines in Fig.~\ref{Fig5}(b)]. 
For $\kappa=-0.5$, $-0.25$ and $0$, $G_e/G_0$ is independent of $\hbar\omega_{x,1}$. This is expected since $G_e/G_0$ is below or close to $0.5$.
Unique symbols are used to represent data from split gates of each length. The dashed lines are linear fits as a guide to the eye \cite{note1}. 
 
In the sub-open regime $G_e/G_0$ reduces with decreasing $\hbar \omega_{x,1}$. This is consistent with our previous work~\cite{Smith2015}. However, in Ref.~\cite{Smith2015} an array of identical length split gates was measured in which differences in the electrostatic profile between devices arose only from variations in the background potential.
In Fig.~\ref{Fig5}(c) data from split gates of different length all follow the same trend line for a given $\kappa$. 
This implies that the profile of the longitudinal barrier--rather than the gate length--is the most significant factor governing the conductance value of the 0.7 anomaly. 

The data in Fig.~\ref{Fig5} are presented in a similar way to Fig. 2 of Ref.~\cite{Smith2015} for ease of comparison.
We emphasize that these are entirely different datasets, from separate devices. The similarity arises because the same technique developed in Ref.~\cite{Smith2015} to analyze the 0.7 anomaly is applied here.

The importance of device-specific confining potential has been highlighted in Ref.~\cite{Burke2012}, which summarizes numerous studies of the 0.7 anomaly as a function of carrier density. The conductance of the 0.7 anomaly has been seen to both increase and decrease as a function of carrier density. These conflicting trends likely arise due to the extreme sensitivity of the 0.7 anomaly to differences in the electrostatic potential between devices~\cite{Burke2012}.

The trend in Fig.~\ref{Fig5}(c) is also consistent with the split-gate length dependence reported in Ref.~\cite{Reilly2001}. The 0.7 anomaly occurred at lower values as gate length increased (three devices $\approx$ 0, 0.5 and 2 $\mu$m long were measured). A stronger link between the longitudinal profile and gate length may exist for devices measured in Ref.~\cite{Reilly2001} because they are fabricated on an undoped heterostructure~\cite{Kane1998}, where the absence of dopants may lead to smaller variations in the background potential.

Calculations of the conductance transition between zero and $G_0$ 
using the inelastic scattering model plus the local density of states enhancement \cite{Sloggett2008, Bauer2013} predict a lowering of the conductance in the sub open regime as $\hbar \omega_x$ decreases. These calculations are shown in Fig. S14(b) in supplementary material of Ref. \cite{Bauer2013}, and are in agreement with our data.
Unfortunately, our data do not allow us to distinguish between theories for the occurrence of the 0.7 anomaly since the same trend is predicted in both the spontaneous spin polarization \cite{Jaksch2006} and Kondo scenarios \cite{Hirose2003} (this is discussed in more detail in Ref. \cite{Smith2015}). However, the inelastic scattering scenario is the only theory for which detailed calculations have been performed as a function of $\hbar \omega_x$.

Figure~\ref{Fig5}(d) shows $G_e/G_0$ as a function of 1D subband spacing $\Delta E_{1,2}$. 
There is a slight reduction of $G_e$ with increasing $\Delta E_{1,2}$
in the sub-open regime, although the trend is weak. 
As seen in Fig. \ref{Fig2}(b), a larger $\Delta E_{1,2}$ occurs for devices with a more negative pinch off voltage. This suggests a possible explanation for the trend in Fig.~\ref{Fig5}(d): the strength of transverse confinement is stronger for the devices with larger $\Delta E_{1,2}$, leading to an increase of strength of electron interactions, thus affecting the conductance value of the 0.7 anomaly.
This may be understood within the framework of the inelastic scattering model, which makes predictions of the effect of electron interaction strength on the conductance of the 0.7 anomaly.
For example, Fig. S14(c) of supplementary material for Ref.~\cite{Bauer2013} shows a reduction of the conductance in the sub-open regime with increasing interaction strength, for a constant barrier curvature. 
As a final point of interest, the trends in Figs.~\ref{Fig5}(c) and \ref{Fig5}(d) do not depend on the split gate length, compatible with the inelastic scattering model which is based purely on the shape of the potential barrier. 

\section{Conclusion}
We have systematically studied the effect of changing split gate size on device behavior. This is achieved using a multiplexing technique which allows many nanostructure devices to be compared on a single cooldown. Multiple devices of each size are measured, providing statistical information on the variance of conductance properties between individual designs.
In total, we measured 95 split gates with 7 different length/width combinations.
Increasing the gate length and/or reducing the width moves the average voltage required to define a 1D channel and the pinch off voltage closer to zero. The 1D subband spacing also reduces for devices with longer gates.
Further, the percentage of devices displaying accurate quantization of conductance reduces dramatically as the area of the channel increases.

The electrostatic environment in the 2DEG is very influential on the 1D potential profile. The significance of the background potential is highlighted by three key results reported here. 
Firstly, the spread in values of the 1D subband spacing and the 1D barrier curvature for a given gate length are as large as the overall variation in the average values of these properties over the range of gate lengths measured.
Secondly, the lever arm $\alpha$ which depends on the coupling between the gate and the 1D channel does not continue to increase as a function of gate length beyond $L=0.7$ $\mu$m, a deviation from the expected result.
Thirdly, the curvature of the longitudinal potential barrier estimated using a saddle-point model is not strongly related to gate length.

These results imply that \emph{i}) gate size cannot be relied upon as a good indicator of the length of the 1D channel, and \emph{ii}) the background disorder potential is at least as significant as gate size in determining the potential landscape in the 1D channel.

The 0.7 anomaly is compared between split gates of different length. The conductance value of the 0.7 anomaly reduces as the barrier curvature becomes shallower, rather than depending specifically on the split gate length. The particular confining potential in each device, and principally the barrier curvature, may be the primary factor governing the conductance of the 0.7 anomaly at a given temperature and magnetic field.

This study has characterized split gates on a modulation doped GaAs/AlGaAs heterostructure.
A potential avenue for future work is to perform measurements on undoped heterostructures where the absence of Si donors results in a reduction of background disorder~\cite{Kane1993}, leading to more reproducible behavior both between split gate devices~\cite{Sarkozy2009, Sfigakis2010} and after thermal cycling~\cite{See2012}. 
Measuring an array of devices on an undoped structure may also give insight on the degree to which disorder affects nanostructures fabricated on undoped heterostructures~\cite{See2015}.

\section{Acknowledgements}
This work was supported by the Engineering and Physical Sciences Research Council Grant No. EP/I014268/1. The dataset for this article is available at www.repository.cam.ac.uk/handle/1810/249174. The authors thank A. R. Hamilton, J. von Delft, S. Ludwig, and E. T. Owen for helpful discussions, and R. D. Hall for e-beam exposure.

$\dagger$ Present address: Department of Physics, University of Wisconsin-Madison, Madison, WI 53706, United States.

$\ddagger$ Department of Electronic $\&$ Electrical Engineering, University of Sheffield, Sheffield, S1 3JD, United Kingdom.

$*$ Corresponding author: luke.smith@wisc.edu


\begin{thebibliography}{MM}
\bibitem{NoteError} The error associated with the mean value is given by the square root of the averaged square of individual errors divided by the number of data points.
\bibitem{NoteA} Data which are included are those which can be fitted with a parabolic potential. This is a self-selecting process, and may explain why the range of $\hbar\omega_{x,1}$ is within that which was found for the lithographically identical devices.
\bibitem{note1} A linear fit is plotted to highlight the upward trend. This is simply a guide to the eye. The range of $\hbar \omega_{x,1}$ is too small to determine the exact nature of the relationship between $G_e$ and $\hbar \omega_{x,1}$.
\end{thebibliography}
\end{document}